\documentclass[a4paper]{paper}

\usepackage{graphicx}
\usepackage[figurename=Fig.]{caption}
\usepackage{hyperref}

\usepackage{amsmath}
\usepackage{amsfonts}
\usepackage{amssymb}

\usepackage{mathtools}
\usepackage{framed}
\usepackage{siunitx}
\usepackage[dvipsnames]{xcolor}
\usepackage{comment}
\usepackage{epstopdf}
\usepackage{float}
\usepackage{layout}
\newfont{\amsbold}{msbm10}
\newfont{\logobold}{logobf10 scaled\magstep2}

\def\*#1{\mathbf{#1}}

\newcommand{\bn}{\mathbf{n}}

\newcommand{\bu}{\mathbf{u}}

\newcommand{\bx}{\boldsymbol{x}}
\newcommand{\by}{\mathbf{y}}

\newcommand{\ddx}[2]{\frac{\partial #1}{\partial #2}}

\newcommand{\mbu}{\mathbf{u}}

\newcommand{\p}{p}

\newcommand{\mbg}{\mathbf{g}}
\newcommand{\mbnull}{\mathbf{0}}

\newcommand{\initialM}{\mathcal{M}}
\newcommand{\gauss}[2]{\mathcal{N}(#1,#2)}
\newcommand{\R}{\mathbb{R}}
\newcommand{\ndof}{n_{\text{dof}}}

\newcommand{\testFunc}{w^h}
\newcommand{\up}{c_{n+1}}
\newcommand{\un}{c_n}
\newcommand{\pp}{p_{n+1}}
\newcommand{\pn}{p_n}
\newcommand{\dt}{\Delta t}
\newcommand\norm[1]{\left\lVert#1\right\rVert}

\DeclareMathOperator*{\argmin}{arg\,min}
\newcommand{\mmap}{m_{\text{map}}}

\usepackage{jabbrv}

\title{Contaminant Dispersion Simulation \break in a Digital Twin
Framework \break for Critical Infrastructure Protection}

\author{Max von Danwitz$^1$, Jacopo Bonari$^1$, Philip Franz$^1$, Lisa Kühn$^1$, Marco Mattuschka$^1$, Alexander Popp$^{1,2}$}

\heading{M. v. Danwitz, J. Bonari, P. Franz, L. K\"{u}hn, M. Mattuschka, A. Popp}

\address{$^{1}$ German Aerospace Center (DLR)\\
Institute for the Protection of Terrestrial Infrastructures\\ Simulation Methods for Digital Twins\\
Rathausallee 12, 53757 Sankt Augustin, Germany\\
email: max.vondanwitz@dlr.de, www.dlr.de/pi
\and
$^{2}$ University of the Bundeswehr Munich\\
Institute for Mathematics and Computer-Based Simulations (IMCS)\\
Werner-Heisenberg-Weg 39, 85577 Neubiberg, Germany\\
email: alexander.popp@unibw.de, www.unibw.de/imcs
}


\abstract{ A digital twin framework for rapid predictions of atmospheric contaminant dispersion is developed to support informed decision making in emergency situations. In an offline preparation phase, the geometry of a built environment is discretized with a finite element (FEM) mesh and a reduced-order model (ROM) of the steady-state incompressible Navier-Stokes equations is constructed for various wind conditions. Subsequently, the ROM provides a fast wind field estimate based on the current wind speed during the online phase. To support crisis management, several methodological building blocks are combined. Automatic FEM meshing of built environments and numerical flow solver capabilities enable fast forward-simulations of contaminant dispersion using the advection-diffusion equation as transport model. Further methods are integrated in the framework to address inverse problems such as contaminant source localization based on sparse concentration measurements. Additionally, the contaminant dispersion model is coupled with a continuum-based pedestrian crowd model to derive fast and safe evacuation routes for people seeking protection during contaminant dispersion emergencies. The interplay of these methods is demonstrated in two critical infrastructure protection (CIP) test cases. Based on simulated real world interaction (measurements, communication), this article demonstrates a full Measurement-Inversion-Prediction-Steering (MIPS) cycle including a Bayesian formulation of the inverse problem.}

\graphicspath{{fig}}

\begin{document}

\renewcommand{\figureautorefname}{Fig.}
\renewcommand{\equationautorefname}{Eq.}
\renewcommand{\sectionautorefname}{Sec.}
\renewcommand{\subsectionautorefname}{Sec.}

\thispagestyle{empty}

\section{Introduction}
\label{sec:intro}

Atmospheric dispersion of hazardous materials poses a significant threat to persons operating and using critical infrastructures. The release of contaminant may occur accidentally in industrial leaks, spills, or explosions~\cite{Patnaik.2012}, or deliberately in an act of sabotage or terrorism~\cite{Boris.2002}. Examples of vulnerable infrastructure components include critical entities of transport infrastructure such as train stations or chemistry plants in the energy sector. In emergency situations, real-time predictions of contaminant dispersion are urgently needed for informed decision-making, e.g., in an evacuation scenario~\cite{Wogrin.2023}. To respond to this need, the digital twinning of a built environment for chemical accident response is proposed in this paper. Outlining the targeted capabilities of the digital twin, we pose the following questions:
\begin{itemize}
    \item How is contaminant transported from a known initial distribution?
    \item Are sparse sensor measurements sufficient to identify the source location?
    \item What is the best path for crowds seeking protection during an emergency evacuation?
\end{itemize} \noindent
While the first question addresses a classical, yet intricate, forward problem of computational fluid dynamics (CFD), the later questions lead to inverse and optimization problems. The inverse problem, in particular, requires a combination of measurement data and physical models following a hybrid digital twinning strategy~\cite{Danwitz.2023b}. The combined use of physics-based and data-driven models is also denoted as hybrid analysis and modeling (HAM)~\cite{San.2021}. Answers to these questions contribute to the overarching goal of constructing a hybrid digital twin that enables the responsive and fast planning of safe evacuation paths under consideration of contaminant dispersion in a built environment.

Computational methods to investigate airborne contaminant transport span a remarkable range in terms of prediction accuracy, computational cost and application customization. Established approaches with (almost) zero latency are often based on a Gaussian dispersion model describing the concentration distribution with a Gaussian distribution parameterized by empirical parameters~\cite{Palazzi.1982}. As an example, the widely-used software package ALOHA (Areal Locations of Hazardous Atmospheres) produces a threat zone estimate where a user-specified level of concern is exceeded~\cite{Jones.2013}. On the one hand, the underlying model accounts for the properties of the considered chemical species, release circumstances, and weather conditions. On the other hand, a flat terrain is assumed and the influence of buildings and topographic features is neglected in the transport description.

In terms of computational cost, the other end of the spectrum is marked by high-fidelity CFD simulations performed on high-performance computing (HPC) architectures. Numerical methods in this field reach from application-driven finite volume methods~\cite{Du.2021}, to versatile finite element methods (FEM)~\cite{Bazilevs.2007} and high-order Discontinuous Galerkin methods with appealing numerical properties such as in-built stabilization mechanisms~\cite{Fehn.2019}. These methods provide a high level of prediction accuracy, yet their computational cost is prohibitive when considering solver-in-the-loop problems in a near real-time setting. To facilitate such applications, this contribution sacrifices some prediction accuracy to reduce computational cost and selects a simple model for contaminant dispersion. The wind field in a built environment is estimated with incompressible Navier-Stokes equations, and contaminant transport is modeled as advection-diffusion process. In terms of model output complexity, the suggested approach is comparable with dispersion nomographs proposed by Boris~\cite{Boris.2002}.

However, the focus of the present contribution is not on the dispersion modelling itself, but rather on the integration of numerical methods in an open-data based digital twin framework for urban physics simulations. To address crisis management topics, such as source detection and emergency evacuation, several methodological building blocks are combined. Automatic FEM meshing~\cite{Geuzaine.2009} and solving capabilities~\cite{FEniCS_book} are extended with reduced-order models (ROM)~\cite{Rozza.2024} to enable fast forward-simulations of contaminant dispersion. Further methods are integrated in the framework to address inverse problems, such as source detection~\cite{Villa.2021} and control mechanisms in evacuation scenarios~\cite{Pietschmann.2024}. The interplay of these methods is demonstrated in two critical infrastructure protection (CIP) test cases.

\section{Digital Twinning and Simulation Workflow}
\label{sec:digitwin}
\begin{figure}
    \centering
    \includegraphics[width=0.35\textwidth]{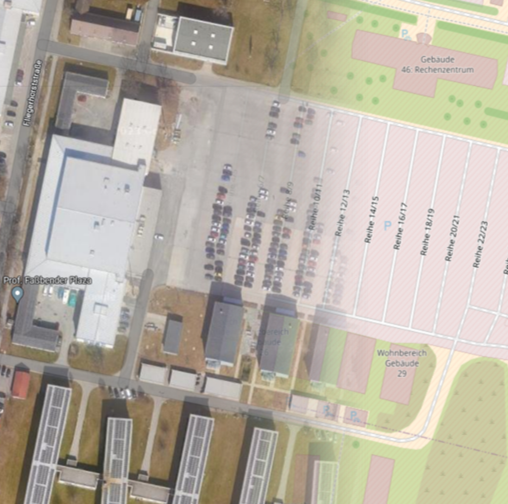}
    \includegraphics[width=0.62\textwidth]{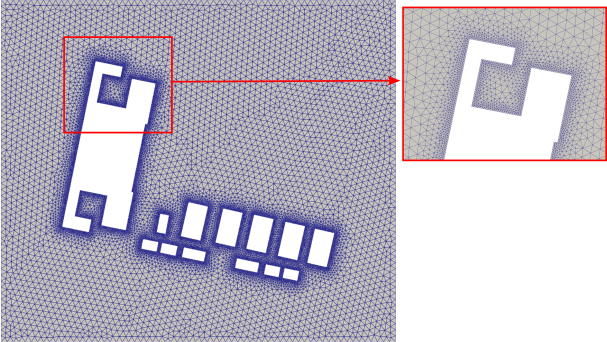}
    \caption{Virtual replica of built environment based on OSM data (left) and FEM mesh (right)}
    \label{fig:digital_twin}
\end{figure}

Following the recently proposed conceptual digital twin framework for infrastructure protection~\cite{brucherseifer_digital_2021}, a few terms are reintroduced in the urban physics simulation context of this paper. The considered \emph{real world section} is a part of the campus of University of the Bundeswehr Munich, employed as a representative built environment. This physical asset is represented digitally by a \emph{virtual replica}. In~\autoref{fig:digital_twin}, the real world section is represented by a satellite image which is overlaid with its virtual replica (based on Open Street Map (OSM) data). For urban physics simulations, the virtual replica has to include a geometry description from which \emph{virtual clones} are instantiated to perform simulations. With real-time data from the physical asset (e.g., wind speed) included in the virtual replica a \emph{digital shadow} is obtained. To form a digital twin, the \emph{bi-directional coupling} of the real asset and its digital representation is required, e.g., in~\autoref{sec:cip}, sensor measurements of the contaminant concentration (in the future) from the physical world are used in the digital twin to compute optimized evacuation paths, which are (planned to be) communicated to evacuation helpers to guide pedestrians in the real physical infrastructure. The interaction with the proposed digital twin can be broadly organized in two phases, an \emph{offline preparation phase} with the construction of the digital twin and an \emph{online action phase}. 
Details of the offline preparation phase were previously presented in~\cite{Bonari.2024}. During the online phase, the digital twin provides added-value to critical infrastructure operators. The constructed digital twin has simulation capabilities of airborne contaminant transport to predict the future evolution of a known contaminant distribution (\autoref{sec:ad}). Moreover, unknown contaminant source locations can be inferred from measurements based on the wind field estimate (\autoref{ssec:inverse}). The following enumeration gives an overview of the complete online simulation workflow for emergency evacuations, with details of the mathematical modelling provided in the following sections.
\begin{enumerate}
\item Measure wind speed $u_{\text{in}}$ and contaminant concentration $d$ at sensor locations.
\item Using $u_{\text{in}}$, query the ROM to obtain a wind field estimate $\bu^*$.
\item Combine wind field $\bu^*$ and measurements $d$ to infer the initial condition $m$.
\item With inferred initial condition, predict concentration evolution $c(t)$ for the time of interest.
\item Using $c(t)$, compute optimized evacuation paths that avoid contaminated areas. 
\end{enumerate}
Based on (currently simulated) real world interaction (measurements, communication), this article demonstrates a full Measurement-Inversion-Prediction-Steering (MIPS) cycle including a Bayesian formulation of the inverse problem~\cite{Wogrin.2023}.

\section{Wind field estimation}
\label{sec:wind}
\begin{figure}
    \centering
    \includegraphics[width=0.45\textwidth]{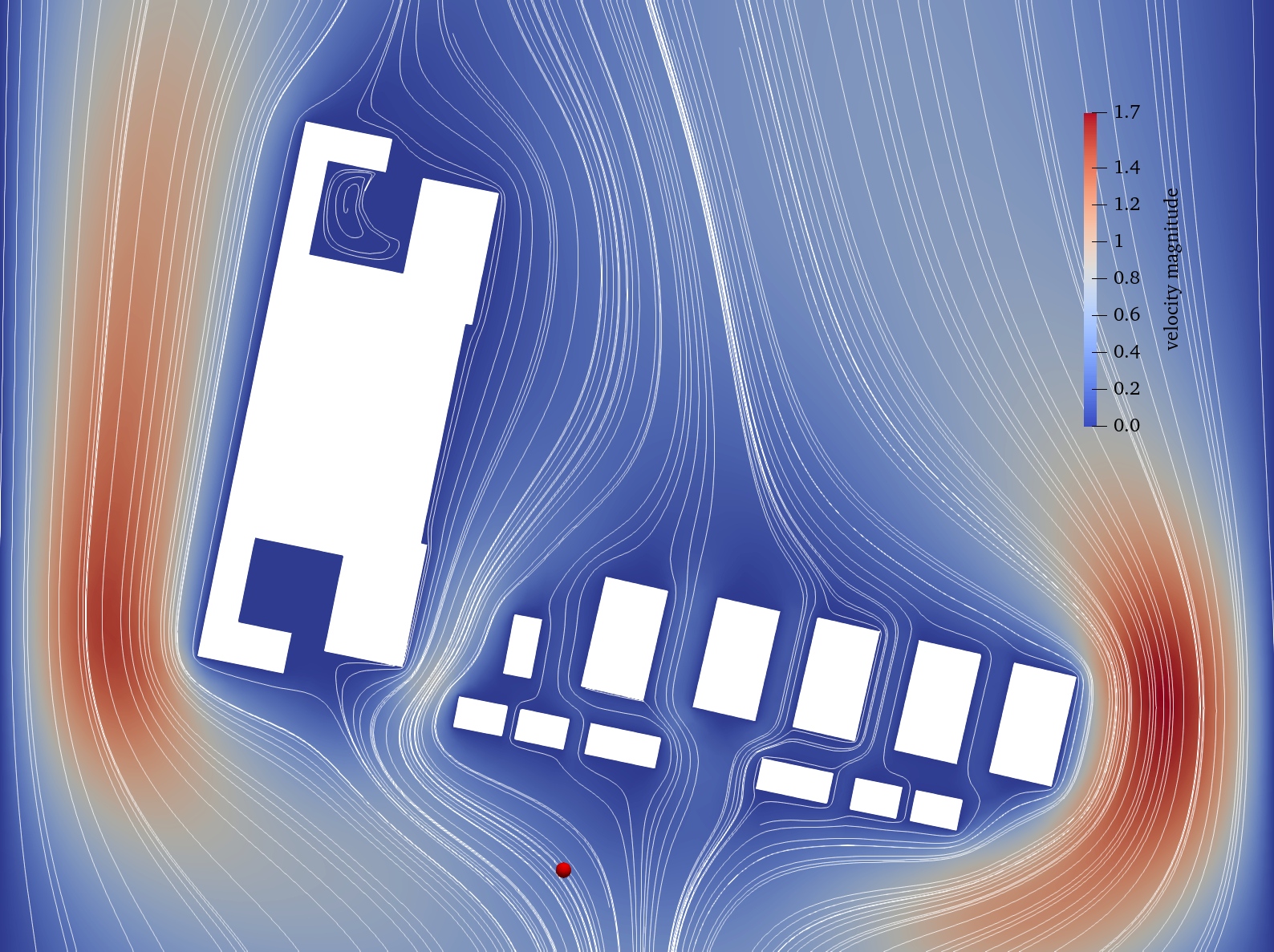} \hspace{0.5cm}
    \includegraphics[width=0.45\textwidth]{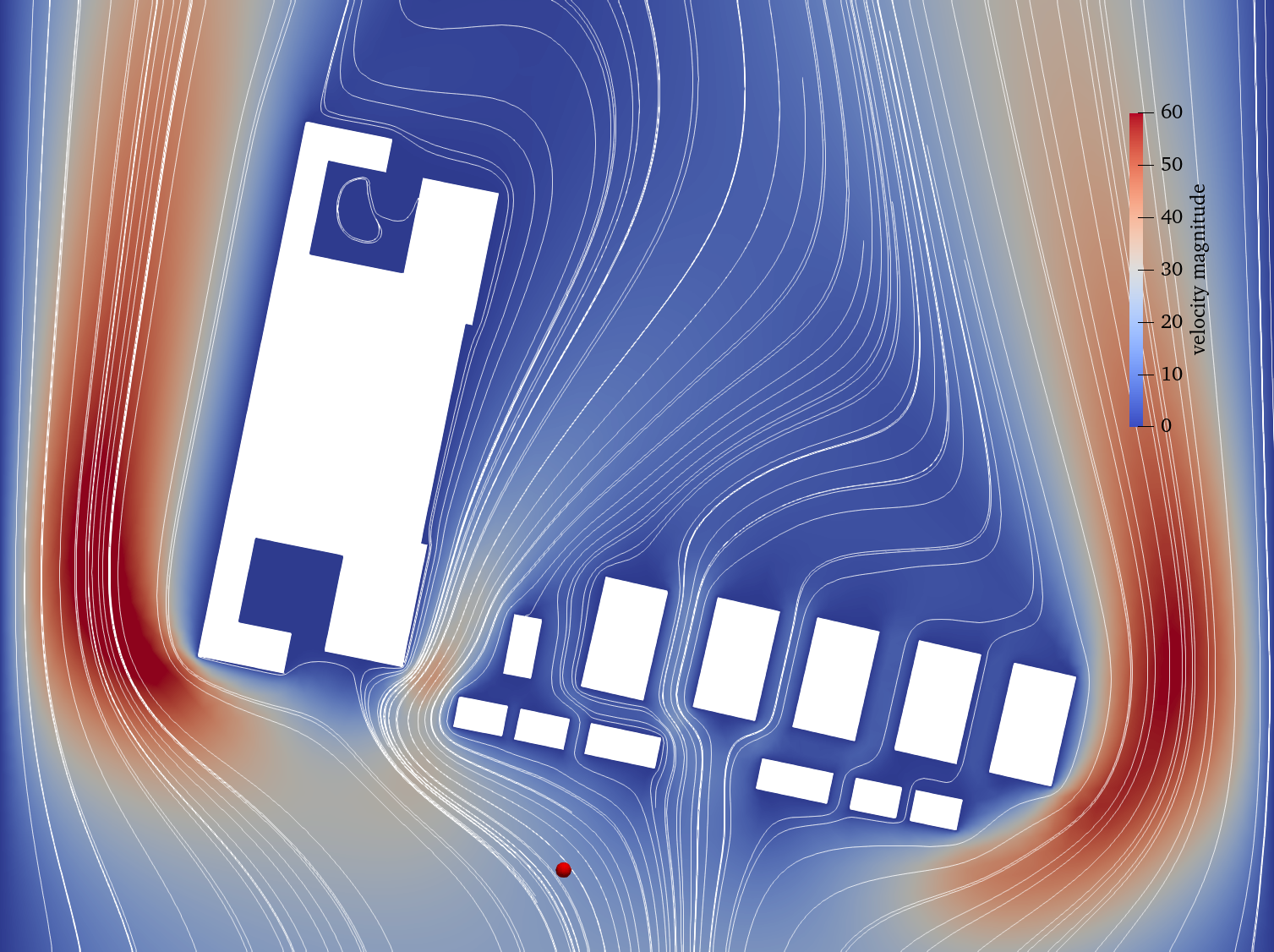}
    \caption{Wind field estimation with ROM approximation of incompressible Navier-Stockes equations for minimal parameter value $u_\text{in}=\SI{0.5}{\meter\per\second}$ and maximal parameter value $u_\text{in}=\SI{20}{\meter\per\second}$}
    \label{fig:ins_fom}
\end{figure}

The goal of the wind field estimation is to compute a spatially resolved flow field in a built environment for given wind direction and wind speed. To facilitate a fast evaluation within a ROM, the underlying physics are described as parametric partial differential equations, namely the steady-state incompressible Navier-Stokes equations~\cite{Elman.2014}. In terms of the flow velocity $\mbu$, and pressure $\p$, they read
\newcommand{\normal}{\bn}
\begin{equation}\label{eq:INS_strong}
    \begin{aligned}
        -\nu \nabla^2\mbu + \mbu\cdot\nabla\mbu + \nabla \p &= \mbnull \quad &\text{in} \quad &\Omega, \\ 
        \nabla \cdot \mbu &= 0 \quad &\text{in} \quad &\Omega, \\
        \mbu &= \mbg(\mu) \quad &\text{on} \quad &\Gamma_D, \\
        \nu \frac{\partial \mbu}{\partial \normal} - \p \normal &= \mbnull \quad &\text{on} \quad &\Gamma_N,
        \end{aligned}
\end{equation}
where $\nu$ denotes the kinematic viscosity, $\normal$ the outwards surface normal vector on $\Gamma_N$ and boundary conditions are defined on the boundaries $\Gamma_D$ and $\Gamma_N$, respectively. Variations of the wind speed are considered as inflow velocity on the Southern domain boundary and, hence, the Dirichlet boundary condition $g$ is parameterized by $\mu \in [\SI{0.5}{\meter\per\second} \ldots \SI{20}{\meter\per\second}]$. In consequence, flow field and pressure distribution depend on the parameter $\mu$ as $\left(\mbu,\p\right) = \left(\mbu(\mu),\p(\mu)\right)$.

\newcommand{\reynolds}{\text{Re}}
A Taylor-Hood finite element discretization of the computational domain facilitates a stable numerical solution of~\autoref{eq:INS_strong}. To enable a fast estimation of the wind field in a hazardous situation, a ROM is constructed based on a proper orthogonal decomposition (POD) of finite element solutions that sample the parameter space (snapshots) and a subsequent Galerkin projection~\cite{Rozza.2024}. With an moderate number of basis functions, a sufficiently accurate approximation is obtained, and the ROM achieves an average speed-up of 50 over the parameter range
~\cite{Bonari.2024}. The Reynolds number $\reynolds$ provides an indicator for the flow behavior based on characteristic flow velocity $u_c$, length $L_c$ and viscosity. If not stated otherwise, the following parameters are employed in the wind field estimation, resulting in a laminar flow approximation with
$   \reynolds_c = \frac{u_c L_c}{\nu} = 10 \, \text{with}\, u_c = u_{in} = \SI{10}{\meter\per\second},\, L_c = \SI{100}{\meter},\, \nu = \SI{100}{\square\meter\per\second}.
$

\begin{figure}
    \centering
    \includegraphics[width=0.6\textwidth]{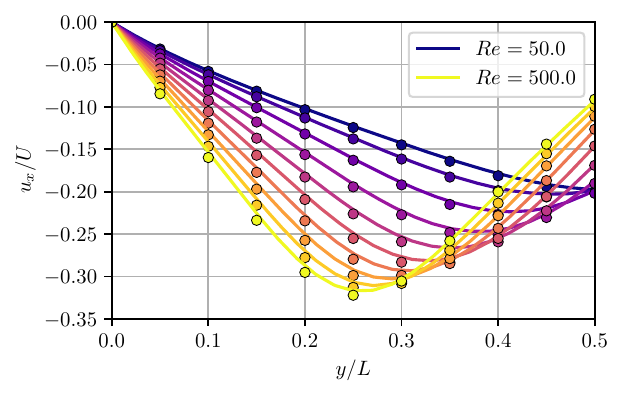}
    \caption{Verification example. Results of the proposed implementation (lines) reproduce the reference solution (circular marks) in the parameter range $\reynolds \in [50,\dots,500]$ very well.}
    \label{fig:lid-driven}
\end{figure}

To verify our implementation based on the open-source software FEniCS~\cite{FEniCS_book}, simulation results of the classic lid-driven cavity benchmark problem~\cite{Elman.2014} are compared to a reference solution of a commercial finite element code. The test problem is defined on a square domain of side $L$, with a tangential driving velocity with magnitude $U$ applied to the top side. In the test, ten linearly spaced values of the Reynolds number in the range $[50,\dots,500]$ have been investigated. Results are shown in Fig.~\ref{fig:lid-driven} in terms of the horizontal velocity $u_x$ along the lower half of a vertical line at $\frac{x}{L}=0.5$. The solid lines represent results coming from our implementation, while the circles mark the solution of the commercial software. For both sets of results, the same indigo color is used for the lowest value of $\reynolds=50.0$, which then fades to yellow approaching the maximum of $\reynolds=500.0$. The agreement of the two solutions demonstrates the robustness of the implementation for the parameter range considered in the following sections.

\section{Contaminant transport simulation}
\label{sec:ad}
The wind-induced airborne transport of released contaminant is modeled with the time-dependent advection-diffusion equation:
\begin{equation}
\text{res}(c) \coloneqq  \ddx{c}{t} + \bu \cdot \nabla c - k \, \Delta c = 0. \label{eq:ADstrong}
\end{equation}
Therein, the scalar unknown $c(\bx, t)$ represents the contaminant concentration as a function of the spatial coordinates $\bx$ and time $t$. The advection velocity is a given vector field $\bu(\bx)$, obtained as solution of Eq.~\eqref{eq:INS_strong} and in practice approximated with the ROM. Furthermore, the constant diffusion coefficient is denoted by $k$. The corresponding initial boundary value problem (IBVP) states that we require Eq.~\eqref{eq:ADstrong} to hold on $\Omega$, along with a known initial concentration distribution $m(\bx)$ and given boundary conditions. The complete IBVP reads:
\begin{align}
\label{eq:IBVP}
\text{IBVP} \quad
    \begin{cases}
    	\text{res}\bigl(c(\bx, t)\bigl) = 0, &\;  \text{in} \;  \Omega \times (0,T),  \\
    	c(\bx, t) = 0, &\;  \text{in} \; \Gamma^\mathrm{D}\times (0,T), \\
    	\partial c/\partial \mathbf{n}= 0, &\;  \text{in} \; \Gamma^\mathrm{N}\times (0,T), \\ 
    	 c(\bx, 0) = m(\bx),& \;\text{in} \; \Omega.   \\
    \end{cases}
\end{align}

With suitable test and trial function spaces $\mathcal{V}_{h,n}$ and $\mathcal{S}_{h,n}$, the well established SUPG-stabilized weak form of the boundary value problem can be stated as follows~\cite{Brooks.1982}:
Find $c^h \in \mathcal{S}_{h,n}$ such that for all $w^h \in \mathcal{V}_{h,n} $
\begin{alignat}{2}
\label{eq:stabilizedAD}
0 =& & &\int \testFunc \cdot \left( \ddx{c^h}{t}  +  \bu \cdot \nabla c^h \right) \, d\Omega + \int \nabla \testFunc \cdot \left( k \nabla c^h \right) \, d\Omega 
+\int \left( \bu \cdot \nabla \testFunc \right) \cdot \tau \cdot \text{res}\left(c^h\right)  \, d\Omega . 
\end{alignat}
To account for local characteristics of the initial boundary value problem, we define the stabilization parameter $\tau= \min \left( \frac{h_K^2}{2k}, \frac{h_K}{\| \bu\| }\right)$
using the element diameter $h_K \coloneqq \sup_{\bx,\by \in K}|\bx -\by|$. For a more advanced metric-based stabilization parameter definition, we refer to~\cite{Danwitz.2023}. The transient nature of the problem is treated with an implicit Euler time stepping algorithm~\cite[Equation (10.25)]{Elman.2014}. With $ \ddx{c^h}{t} \approx \frac{\up - \un}{\dt}$ and $u_{n=0} = u_0$, the following discretized weak form is obtained:
\begin{alignat}{2}
	\label{eq:weakAD}
	& & & \int \testFunc \cdot \up \, d\Omega 
	+ \dt \int \testFunc \cdot \left( \bu \cdot \nabla \up \right) \, d\Omega  
	+ \dt \int \nabla \testFunc \cdot \left( k \nabla \up \right) \, d\Omega \\ \nonumber
	&+& & \int \bu \cdot \nabla \testFunc \cdot \tau \cdot \up \, d\Omega 
	+ \dt \int \bu \cdot \nabla \testFunc \cdot \tau \cdot \left( \bu \cdot \nabla \up - \nabla \cdot k \nabla \up  \right) \, d\Omega \\ \nonumber
	&=& & \int \testFunc \cdot \un \, d\Omega + \int \bu \cdot \nabla \testFunc \cdot \tau \cdot \un \, d\Omega.
\end{alignat}
After an initial assembly of the resulting linear system of equations, time-stepping only requires an update of the right hand side and the solution of the system. This computational setup is particularly well-suited for the frequent evaluation of the forward problem with varying initial conditions, which is required in the solution process of inverse problems.
 \begin{figure}[t]
    \centering
    \includegraphics[width=1.0\textwidth]{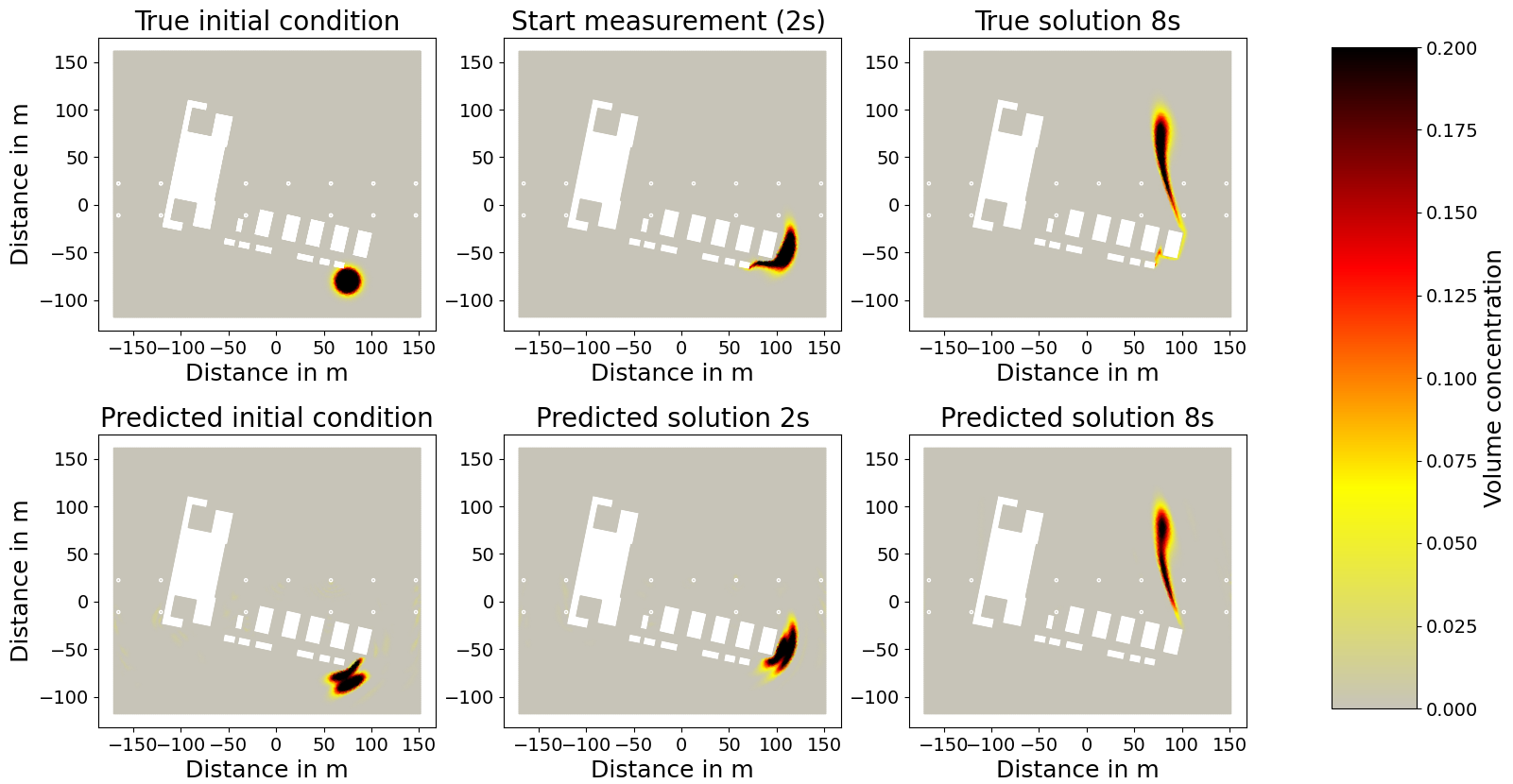}
    \caption{Inverse problem solution for measurements at 13 sensor locations starting after 2 seconds}
    \label{fig:inverse_2d_pred}
\end{figure}

\section{Critical Infrastructure Protection Application Cases}
\label{sec:cip}

\subsection{Inverse problem: Contaminant source identification}
\label{ssec:inverse}

\noindent
 A description of the transport of contaminated air in a bounded open domain $\Omega \subseteq \mathbb{R}^n \text{ for } n\in\{2,3\} $ is given in \autoref{eq:IBVP}. The solution $c$ of~\autoref{eq:IBVP} depends on the given geometry $\Omega$, the constant diffusion coefficient $k$, the wind vector field $\bu$ and the initial condition $m \in \initialM$. As the function space for the initial condition, we consider a closed subset of square integrable functions, i.e. $\initialM \subseteq L^2(\Omega)$. An obvious question is whether the initial condition can be reconstructed on the basis of the given measurements. In this setting, the estimation of the initial condition leads to a linear optimization problem. Synthetic measurements are generated in a simulation with a clipped exponential function $ m(x)=\max\{0.5,\exp(-\kappa \norm{x-x_0}_2^2))\}$ as initial condition. 

The center is set to $x_0 = (75, -80)$, and $\kappa$ is chosen, such that the initial condition has notable values only within a radius of \SI{25}{\meter}. The initial condition and results of the simulation are shown in the upper row of~\autoref{fig:inverse_2d_pred}. The concentration evolution shows the expected behavior according to the estimated wind field, compare~\autoref{fig:ins_fom}. 

Sensor measurements are described with the observation operator $\mathcal{B}:C^0({\Omega})\rightarrow \R^{N_s} $, with $N_s$ representing the product of the number of sensors and the number of measurement times. The positions of the 13 sensors are indicated in \autoref{fig:inverse_2d_pred} with white marks. Measurements start after \SI{2}{\second} at a rate of \SI{5}{\per\second}. In addition, measurement uncertainties are modelled with additive noise $\epsilon \thicksim \gauss{0, \Gamma_{\text{noise}} }$. Moreover, it is convenient to introduce the parameter-to-observable map $\mathcal{F}:\initialM \rightarrow \R^{N_s}$ for \autoref{eq:IBVP} as:
\begin{equation*}
\begin{aligned}
\mathcal{F}(m):=\mathcal{B}(c), \text{ s.t. } \text{IBVP}(c,m) \, \text{is fulfilled}.
\end{aligned}
\end{equation*}
Finally, the measurement data can be written as $d=\mathcal{F}(m) + \epsilon$. 

Now, the inverse problem consists of inferring the values of the unknown parameter field $m$ using available measurements $d$. Unfortunately, the problem is heavily under-determined, as sensor measurements are available only at a few locations, yet the initial condition must be reconstructed for the entire domain $\Omega$. For a finite element discretization (\autoref{eq:stabilizedAD}) of the IBVP, the initial condition can be described as a vector $m \in \R^{\ndof}$, where $\ndof$ is the number of degrees of freedom in~\autoref{eq:weakAD}, with $\ndof \gg N_s$. In order to transform this into a well-posed problem, prior knowledge is needed, and we formulate the inverse problem in a Bayesian framework~\cite{Villa.2021}. A Gaussian prior $\gauss{m_{\text{pr}},\Gamma_{\text{pr}}}$ with mean $m_{\text{pr}}$ and covariance $\Gamma_{\text{pr}}$ is used for the parameter $m$. 
Bayes' theorem for the inverse problem reads 
\begin{equation*}
\begin{aligned}
\pi_{\text{post}}(m|d) \propto \pi_{\text{like}}(d|m) \pi_{\text{prior}}(m)\,.
\end{aligned}
\end{equation*}
Therein, the likelihood function is $\pi_{\text{like}}(d|m) \propto \exp( \frac{1}{2} \norm{\mathcal{F}(m)-d}_{\Gamma^{-1}_\text{noise}}^2)$, and due to linearity of the forward model~$\mathcal{F}$ the posterior distribution is again a Gaussian distribution $ \gauss{\mmap,\Gamma_{\text{post}}}$.
Moreover, the mean value $\mmap$ (that corresponds to the maximum a-posteriori probability) is a reliable estimate for the initial value and thus represents the solution to the inverse problem. To characterize the parameters of the posterior distribution, namely the covariance
$\Gamma_{\text{post}} =  (\mathcal{F}^*\Gamma_{\text{noise}}^{-1}\mathcal{F} - \Gamma_{\text{pr}}^{-1})^{-1}$
and mean $\mmap=\Gamma_{\text{post}}(\mathcal{F}^*\Gamma_{\text{noise}}^{-1}d+\Gamma_{\text{pr}}m_{\text{pr}})\,$,
the formally adjoint operator $\mathcal{F}^*:\R^{N_s} \rightarrow \initialM$ is required. Alternatively, $\mmap$ can be characterized as solution of the minimization problem
\begin{equation}\label{eq:objective}
\begin{aligned}
  \mmap=\argmin_{m\in \initialM} J(m):= \frac{1}{2} \norm{\mathcal{F}(m)-d}_{\Gamma^{-1}_\text{noise}}^2 + \frac{1}{2}\norm{m-m_{\text{pr}}}_{\Gamma^{-1}_{\text{prior}}}^2\,,
\end{aligned}
\end{equation}
with the prior information encoded as a Tikhonov regularization term. Sensor noise at the different locations is assumed to be uncorrelated and of equal magnitude, hence,  $\Gamma_{\text{noise}} = \text{diag}(\sigma^2, \dots, \sigma^2)$. Consequently, the sensor misfit in $J$ can be expressed as $
\norm{\mathcal{F}(m)-d}_{\Gamma^{-1}_\text{noise}}^2 = \frac{1}{\sigma^2}\sum_{i=1}^{N_s}\int_{0}^{T}(\mathcal{B}(c)-d_i)^2\delta_i\,dt\,.$

Choosing an elliptic PDE operator of trace class for the prior covariance $\mathcal{A}=(-\gamma \Delta + \delta I)^{-2}$, where $\gamma>0$ and $\delta > 0$ are parameters of the prior distribution, the inverse problem is well posed~\cite{Villa.2021}. Using the standard Lagrange formalism and calculus of variations~\cite{Troltzsch.2010}, the necessary condition for the parameter $\mmap$ is given by
\begin{alignat}{3}\label{eq:necessary_condition}
   &\mathcal{A}^2(\mmap-m_{\text{pr}}) &&= \mathcal{P}(\mmap)(\cdot, 0) \qquad&&\text{in}\ \Omega\\
   &\gamma \, \nabla \mmap \cdot \eta + \beta \, \mmap&&=0 \qquad &&\text{in}\ \partial \Omega\,.\nonumber
\end{alignat}
with $\mathcal{P}(m)$ being the parameter-to-observable map operator for the associated adjoint mapping to the measured values. This means that $P(m)=p$ and $p$ solves the following adjoint equation for the sensor misfit $f:=\frac{1}{\sigma^2}\,\mathcal{B}^*\sum_{i=1}^{N_s}(\mathcal{F}(m)-d_i)\delta_i$
\begin{equation}\label{eq:strong_adjoint}
\begin{aligned}
  -p_t-k\Delta p - \text{div}(p\textbf{u}) &= -f &\qquad&\text{in}\ \Omega \times (0,T),\\
  (\textbf{u}p+k\nabla p) \cdot \eta &= 0 &&\text{in}\ \Gamma_N  \times (0,T),\\
  p&= 0 &&\text{in}\ \Gamma_D \times (0,T),\\
  p(\cdot, T) &= 0 &&\text{in}\ \Omega.
\end{aligned}
\end{equation}
To solve this numerically, a weak form of \autoref{eq:strong_adjoint} is needed. Due to advection dominance, stabilization techniques of the forward problem are included as in~\autoref{eq:stabilizedAD}. 
Similar to \autoref{eq:weakAD}, the discretization in time for $p^h$ and partial integration yields the equation
\begin{alignat}{2}
	& & & \int \testFunc \cdot \pp \, d\Omega 
	- \dt \int \pp \cdot \left( \bu \cdot \nabla \testFunc \right) \, d\Omega  
	- \dt \int \nabla \pp \cdot \left( k \nabla \testFunc \right) \, d\Omega \\ \nonumber
	&+& & \int \bu \cdot \nabla \pp \cdot \tau \cdot \testFunc \, d\Omega 
	- \dt \int \bu \cdot \nabla \pp \cdot \tau \cdot \left( \bu \cdot \nabla \testFunc - \nabla \cdot k \nabla \testFunc  \right) \, d\Omega \\ \nonumber
	&=& & \int (\pn-f) \cdot \testFunc \, d\Omega + \int \bu \cdot \nabla \pn \cdot \tau \cdot \testFunc \, d\Omega,
\end{alignat}
which is implemented to solve for $\mathcal{P}(\mmap)(\cdot, 0)$. Note that this is a final value problem with $p_{n=0} = p_T$ and time stepping is performed backwards.\\
Again following the approach of \cite{Villa.2021}, the system in \autoref{eq:necessary_condition} can be solved by an inexact Newton conjugate gradient (CG) method using $\mathcal{A}^2$ as preconditioner based on a low-rank approximation of the Hessian $H:=\mathcal{F}^*\Gamma_{\text{noise}}^{-1}\mathcal{F}$. The result of the inverse problem $\mmap$ is denoted as ``predicted initial condition" in the lower row of~\autoref{fig:inverse_2d_pred}. Comparing the upper row (forward simulation) and lower row (based on inverse problem solution) of~\autoref{fig:inverse_2d_pred} illustrates that the inverse problem solution enables the contaminant evolution prediction with satisfactory accuracy.

\begin{figure}[t]
    \centering
    \includegraphics[width=\textwidth]{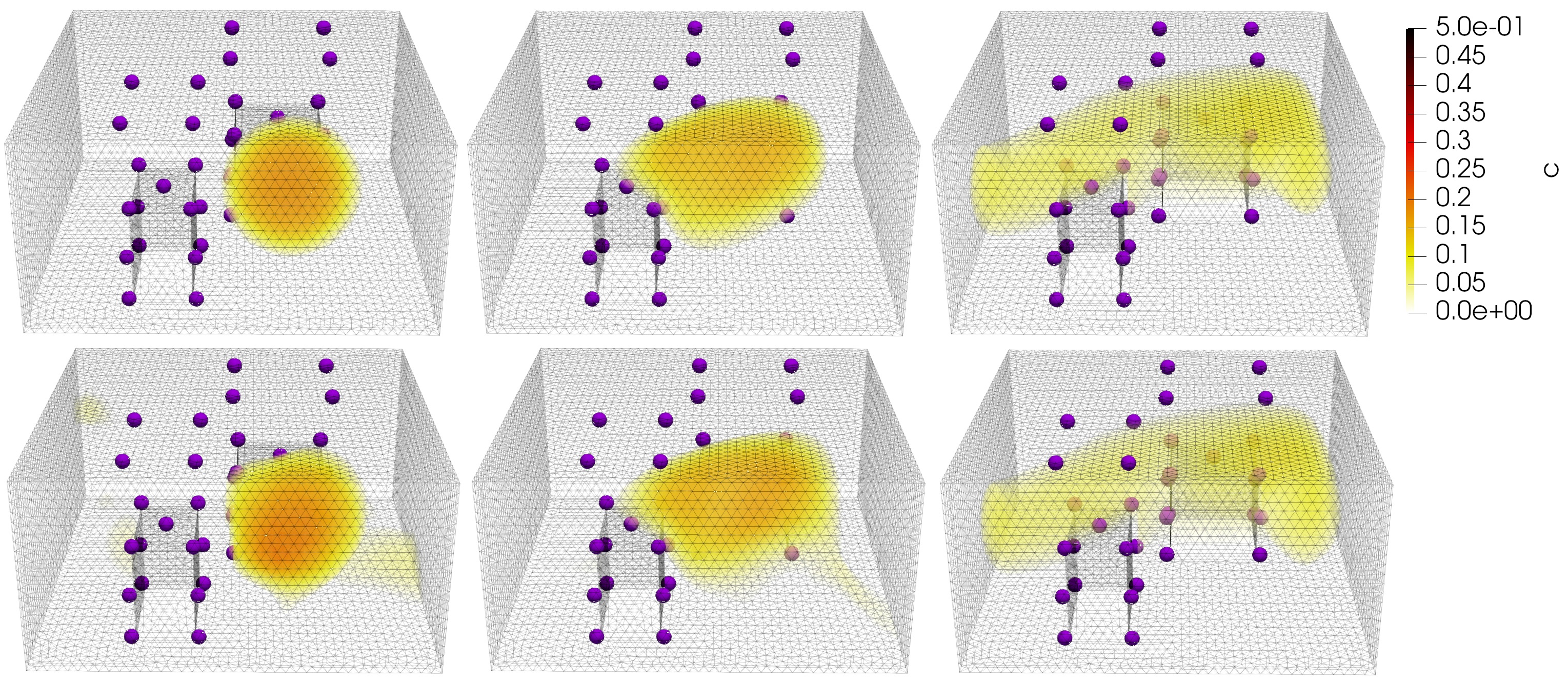}
    \caption{3D solution (top) and prediction (bottom) for t=0.0s, t=1.0s and t=3.0s, where violet
dots mark the 36 sensor positions. More than 300 PDE solutions are needed.
}
    \label{fig:inverse_3d_pred}
\end{figure}\noindent
To further test the applicability of the inverse problem solver, a contaminant transport simulation on a 3D geometry is performed. The experimental setup follows~\cite{Alexanderian.2014} with the wind vector field characterized by $\reynolds = 50$. 
A forward simulation is performed to provide measurement data at the sensor locations (upper row of~\autoref{fig:inverse_3d_pred}). A total of 26 sensors were placed on and around the buildings and, in addition, 5 sensors were placed on virtual ``antennae" above each building. 
The solution of the inverse problem again recovers the true initial condition with acceptable accuracy as shown in \autoref{fig:inverse_3d_pred}. While the methods are easily extended to the 3D geometry, the computational complexity increases significantly.

\subsection{Informed emergency evacuation after chemical accident}
\label{ssec:evac}
\begin{figure}
    \centering
     \includegraphics[width=\textwidth]{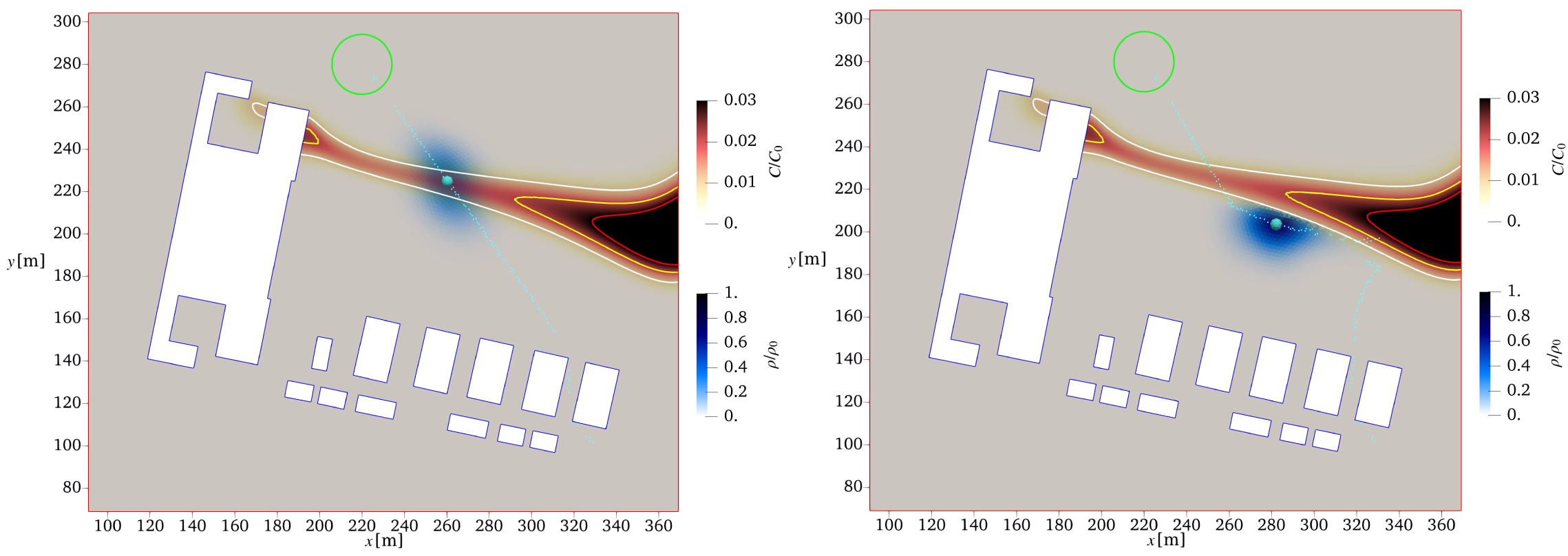}
    \caption{Simulation of pedestrian flow with continuum model (blue). Unaware crowds enter the contaminated area (left), whereas situational awareness enables crowds to avoid the currently contaminated area on their path to the meeting point (right).}
    \label{fig:pedestrian}
\end{figure}

In case of a chemical accident, it is likely that buildings have to be evacuated and people seeking protection gather in designated areas. To provide an example of such an evacuation scenario, we simulate the human motion behaviour based on Hughes continuum model of pedestrian flow~\cite{hughes_continuum_2002}. However, the question arises whether the pedestrians' path to the meeting point is safe in the light of airborne contaminant dispersion after an accident (see~\autoref{fig:pedestrian}, left). In addition, recent simulation approaches allow for introducing informed evacuation helpers in the pedestrian flow model~\cite{Pietschmann.2024}. Evacuation helpers are found to provide a better crowd binding, i.e., pedestrians stay tightly together, which might be advantageous in an evacuation scenario. Based on the proposed digital twin framework, also informed guides can be envisioned to have an improved situational awareness and consider the evolution of the contaminant concentration in the path planning and, hence, enable a safer evacuation. Computationally, the preferred path is found by penalizing the achievable walking speed in a contaminated area. \autoref{fig:pedestrian} provides snapshots of a corresponding pedestrian flow simulation. Note, that the pedestrians are indeed avoiding the currently contaminated area in the constrained solution shown on the right.

\section{Conclusion and Outlook}
\label{sec:conclusion}
Numerical methods to simulate the airborne contaminant transport after a chemical accident were successfully integrated into a digital twin framework. The combination of methods enables the digital twin user to infer the initial contaminant distribution from measurement data and, hence, supports the search for the source location of the contaminant. Provided an improved computational efficiency, the inferred source location can be used to predict the future evolution of the contaminant plume for an informed emergency response. The additional coupling with continuum-based pedestrian flow simulations facilitates the responsive planning of the best paths (fast and free of contaminant) along which evacuation helpers guide people seeking protection towards safe meeting points. The investigation of the presented framework revealed the need to improve the wind field estimation in a built environment. The simultaneous speed-up and increase of accuracy of the wind field estimation while considering the actual atmospheric conditions during the accident constitutes a major future challenge in urban physics research. In further development of the framework, the import of real contaminant measurement data from physical sensors is planned~\cite{Sporrer.2023}, while the visualization of results in a VR-based situational awareness screen can enable the communication with evacuation helpers in the field~\cite{Franke.2023}. In addition, the presented framework can be used to study what-if scenarios and serve as a basis for resilience analysis of critical infrastructure~\cite{Mentges.2023}.

\bibliographystyle{jabbrv_elsarticle-num}
\bibliography{bibliography}

\end{document}